# MicroTCA implementation of synchronous Ethernet-Based DAQ systems for large scale experiments

C. Girerd, D. Autiero, B. Carlus, S. Gardien, J. Marteau, W Tromeur
IPNL, Université de Lyon, Université Lyon I, CNRS/IN2P3, 4 rue Enrico Fermi,
69622 Villeurbanne Cedex

*Abstract*– Large LAr TPCs are among the most powerful detectors to address open problems in particle and astro-particle physics, such as CP violation in leptonic sector, neutrino properties and their astrophysical implications, proton decay search etc. The scale of such detector implies severe constraints on their readout and DAQ system. In this article we describe a data acquisition scheme for this new generation of large detectors. The main challenge is to propose a scalable and easy to use solution able to manage a large number of channels at the lowest cost. It is interesting to note that these constraints are very similar to those existing in Network Telecommunication Industry. We propose to study how emerging technologies like ATCA and µTCA could be used in neutrino experiments. We describe the design of an Advanced Mezzanine Board (AMC) including 32 ADC channels. This board receives 32 analogical channels at the front panel and sends the formatted data through the µTCA backplane using a Gigabit Ethernet link. The gigabit switch of the MCH is used to centralize and to send the data to the event building computer. The core of this card is a FPGA (ARIA-GX from ALTERA) including the whole system except the memories. A hardware accelerator has been implemented using a NIOS II µP and a Gigabit MAC IP. Obviously, in order to be able to reconstruct the tracks from the events a time synchronisation system is mandatory. We decided to implement the IEEE1588 standard also called Precision Timing Protocol, another emerging and promising technology in Telecommunication Industry. In this article we describe a Gigabit PTP implementation using the recovered clock of the gigabit link. By doing so the drift is directly cancelled and the PTP will be used only to evaluate and to correct the offset.

## I. INTRODUCTION

Liquid Argon Time Projection Chambers allow high quality 3D reconstruction of particle trajectories in high volume of detection, perfectly adapted for neutrino experiments. They take the form of a tank filled with liquid Argon maintained at about 87 °K by a cryogenic system. An electric field forces the charges, created by the crossing particle, to drift towards wire chambers located perpendicularly at each the sides of the imaging volume. The charge measurement on each wire allows the calculation of the track coordinates in 2 dimensions. The third dimension is given by the measurement of the arrival time which is proportional to the distance of the interaction from the wire chambers. The initial time is given by the scintillation light produced at the beginning of the interaction and detected immediately by photo-multipliers. For this type of detector the DAQ system should be continuously sensitive. A data streaming and triggerless scheme is well adapted. Zero suppression is applied by overwriting continuously ADC samples as long as there are below the threshold setting. The event building is completely software. Proposals of such very large scale Liquid Argon Detector foresee the use of Liquefied Natural Gas (LNG) Tanks of several ten meters of diameter. In this paper we describe the design of a scalable data acquisition architecture adapted for this new generation of Liquid Argon Time Projection Chambers.

## II. DAQ ARCHITECTURE FOR NEUTRINO EXPERIMENTS

To address this kind of acquisition requirements, classical architecture commonly used in high energy physics experiments are not optimal. Generally, they require powerful readout systems associated with fast trigger processing in order to reduce the amount of data transferred to event building computers. This often leads to complex electronics specific to each experiment. For neutrino experiments, characterized by a rather low data rate, it is possible to avoid this complexity by adopting distributed network architecture associated with a triggerless scheme. This architecture leads to a more optimized electronics with more flexibility for data filtering and event building at software level. The Ethernet network has the advantage to be very wide-spread and in constant technologic evolution. A lot of standard and high performance products are available on the shelf. Despite of this, Ethernet is generally limited to the higher level of data acquisition system and dedicated electronics is preferred near the front-end. The main arguments that have limited the use of Ethernet in acquisition system are his indeterminism and asynchronous characteristics and the low efficiency of network stack.

Ethernet network data acquisition architecture has already been successfully implemented for the OPERA [1] experiment and has proven his efficiency by the simplicity of installation and flexibility of maintenance. The key idea is the design of a Network capable front-end, closed to the sensor, with an embedded micro-processor that ensures all the functionalities required to control and read the detector. This leads to compact acquisition systems. No hardware trigger is generated and no intermediate electronics is necessary. The trigger is decided in software by filtering the arrival data on-line. Data are directly sent to event building computers. The embedded processor on each sensor can be used to apply simple parallel processing on the data such as noise histogram and pedestals.

Only the results are sent to the event building computers. This type of architecture is well adapted for experiments where sensors are distributed. But the drawbacks of such fully distributed implementation are the custom form factor of the sensor boards, the parallel synchronization bus that is necessary for time stamping and the power supply distribution. Several recent and major technological evolutions will definitively cancel the arguments against Ethernet and probably allow the generalization of Ethernet network architecture in data acquisition system. First of all, the Ethernet data rate is now achieving the 100G. This very high bandwidth will allow the concentration of a larger number of channels. The second major event is the creation of the Advanced Telecommunication Computing Architecture (ATCA) [2] and (µTCA) [3] based on high speed serial links among which the 100G and 40G Ethernet. The third argument is the extension of the IEEE1588 standard [4] that allows high accuracy synchronization over the network. The last point concerns the FPGA that includes serializer/deserializer (SERDES) devices able to handle high speed serial links such as 10G.

### III. µTCA FOR NEUTRINO PHYSIC EXPERIMENT

µTCA, gives the possibility to apply this network distributed concept while offering a standard form factor and simplifying the power supply, the cooling, the management and the clock synchronization. The µTCA backplane is based on high speed serial links arranged in single or dual star network topology. Lanes on a MicroTCA backplane are differential high speed Serializer / Deserilizer interconnects (SerDes), with bandwidth capability of at least 3.125 Gb/s supporting a large variety of protocol like for example Ethernet 1G or 10G, PCI Express or SRIO. The boards plugged into a µTCA shelf are called Advanced Mezzanine Card (AMC) [5]. Each AMC board is connected to one or two MicroTCA Carrier Hub through the backplane serial links which provides a central switch function allowing each AMC to communicate with each other or towards external systems through an uplink access. The backplane also provides the connectivity for the clock distribution allowing the AMC board synchronization.

Each AMC includes a Module Management Controller which allows the µTCA carrier management controller to control each AMC board through the IPMB bus. The E-keying is one of the most important features allowing the µTCA carrier to automatically identify the AMC capability and to enable the serial links accordingly. The high speed capability and the modularity of this concept are perfectly adapted for data acquisition in Physics. The only drawback of µTCA is the absence of rear connectivity which only exists in ATCA and that forces to connect the external signals through the front-panel.

A µTCA based architecture is depicted on fig. 1. The idea is to develop an AMC board as a network capable device able to dialog directly with the event building and control computer by a standard Ethernet network. Each AMC board is connected to the MCH Gigabit switch through the backplane lane on port[0].

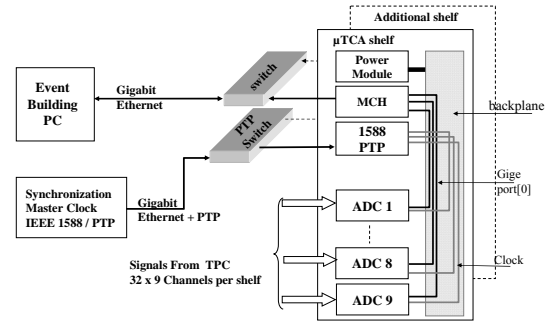

Fig. 1. µTCA based network acquisition system

We have designed an AMC board including 32 ADC channels with analogical inputs at the front panel. The synchronization and the clock distribution are made through the backplane using the IEEE1588 standard also called the precision timing protocol (PTP). The dual star topology of the µTCA shelf allows to separate the synchronization and the readout network. To implement this system we decided to use a µTCA shelf from SHROFF (Fig. 2), able to receive 12 single width and double height AdvancedMC.

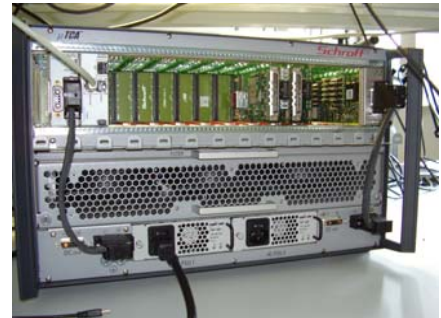

Fig. 2. 12 slots µTCA shelf from SHROFF.

The MCH (UTC001) and the power module (UTC010) are from VADATECH. The Module Carrier Manager Controller manages the power modules, the two CU (Cooling Units) and the 12 AMCs within the µTCA chassis. It also interfaces to and manages the on-board fabric interfaces. The module is available with PCIe, SRIO, 10GbE layer three managed, GbE layer two managed and SAS fabric interfaces. The UTC001 runs Linux 2.6 on its MCMC CPU and is hot-swappable and fully redundant when used in conjunction with a second instance of the module. The firmware is HPM.1 compliant which allows for ease of upgrade.

## IV. AMC ARCHITECUTRE

### A. ADC Mezzanine

The analogical signals from the TPC are connected to the AMC front-panel through a 68 pins connector mounted on a mezzanine board which also includes the ADCs. The AD9212 ADC from Analog Device has been chosen. It includes 8 ADCs with 10 bits resolution and a serial LVDS output. A detailed view of one channel is given on Fig. 3. An AD8138 amplifier is used to make a single ended to differential translation, to cancel the input signal offset and to shift the DC level to the middle range of the ADC inputs.

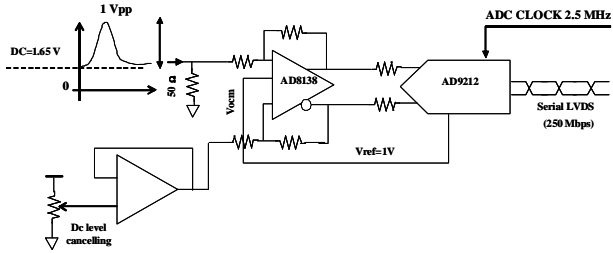

Fig. 3. ADC mezzanine board detail of one channel

This leads in highly integrated design as shown in Fig. 4. We can see the 4 ADC with 8 channels each.

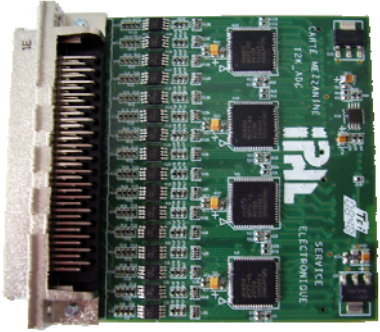

Fig. 4 Analog mezzanine board with 32 ADC channels

This analogical mezzanine board is plugged on the AMC board, using two high Speed Mezzanine connectors (MicroSpeed from ERNI).

### B. AMC mother board

The general bloc diagram of the AMC board is given on Fig. 5. The main component of the AMC board is the FPGA. We decided to use the low cost version of the STRATIX FPGA which is the ARIA-GX. As the Stratix GX it includes SERDES that can be used for Gigabit Ethernet. We implemented a soft-core processor (NIOS II) from ALTERA inside the FPGA in order to reduce the cost and the market dependence from manufacturers. The Medium Access Controller (MAC) is a VHDL IP (GMACII) from the firm Ingenieurbüro Für Ic-Technologie (IFI). This IP includes a checksum generation on the fly, DMA controllers and uses pipelining on both ends. It is perfectly adapted for hardware accelerator implementations. In this design the data are sent using the UDP protocol. Packet headers are built and initialized directly in SDRAM without using any API functions. The UDP data (ADC samples) are directly transfer by DMA between the DPRAM memory and the GMACII buffer. This implementation allows achieving data throughput of 114 Mbytes/s. In order to interface the GMACII IP with GBX ALTERA transceiver through the GMII bus we used the Triple Speed Ethernet MAC from ALTERA with the core variation "1000BASE-X/SGMII PCS only" and the transceiver bloc GXB enabled. It allows a direct connection with the µTCA backplane.

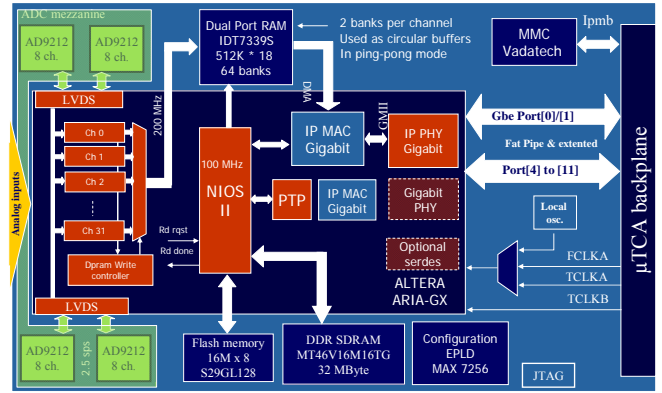

Fig. 5. 32 ADC Advanced Mezzanine Card

The TXP and RXP pins are configured in PCML 1.5 V. The PCS control interface is configured in 1000BASE-X without auto-negotiation. An external switchable bank dual port ram from IDT (72V7339) is used for the data buffer. This memory includes 64 independent 8K x 18 bits banks. The NIOS µP also requires an external SDRAM. A configuration EPLD associated with a Flash memory is used for the configuration management. A simple server running on the NIOS is able to interpret some commands sent by the Event Building. These commands allow to start the ADC sampling, to set the channel threshold and some acquisition parameters.

The NIOS read continuously a bank status register that represents each ADC channel. If a bank contains a valid event the corresponding channel flag is set and the NIOS starts the DMA transfer directly to the MAC.

Each ADC channel is controlled by a channel receiver depicted in Fig. 6. A LVDS receiver deserializes ADC data and stores them into an input FIFO. In parallel the samples are used to detect the trigger condition based on the comparison with the threshold. The write state machine is responsible of writing the ADC into the dual port memory. The dual port memory has 64 banks of 8192x18 words. Two banks (B0/B1) are attached for each channel and work in ping pong mode. The write state machine manages the bank address working as a circular buffer. If no trigger conditions is detected the samples are written continuously into the memory, the oldest

ADC being overwritten by the newest ones. The ADC values and the address location in the bank are written into an intermediate FIFO for each bank (B0/B1). A fifo readout module common to the 32 channels is responsible of reading theses FIFOs and writing the data of the corresponding banks into the dual port memory. This module is based on a pipeline architecture that allows speeding up the FIFO read and write cycles, into the memory. The fifo empty flags, the fifo read enable are pipelined and used to generate directly the memory write enable allowing to read a data from the fifo and to write it in the memory at each clock cycle.

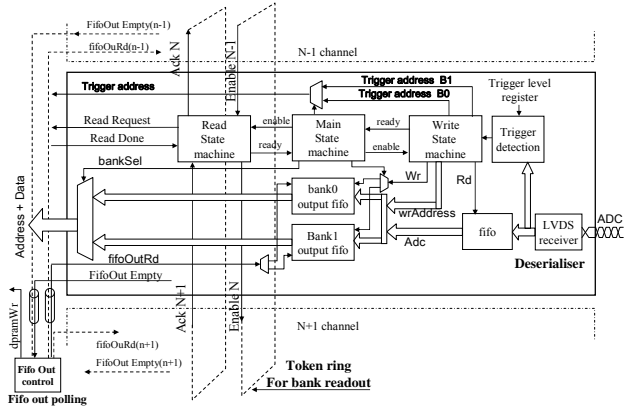

Fig. 6. Detailed bloc diagram of one channel receiver

The memory clock must be at least 32 times faster than the ADC clock period. For an ADC clock of 2.5 MHz, the memory should work at a minimum frequency of 80 MHz. The address location is automatically generated by a combination of the fifo output that contains the address inside the bank and the channel number that gives the bank number. If a trigger occurs, the write state machine continues to store the event corresponding to the TPC drift time keeping just some samples before the trigger. Then, if the second bank is available, it means that this bank has been read, the main state machine tells the write state machine to continue to store the samples into this new active bank. The read state machine sets a flag to advertise that an event is available in a bank. All the read state machines are interconnected through a token ring like structure. The token is passed to next channel once the bank read has been acknowledged by the NIOS. If a channel has no bank available for reading it passes the token to the next channel and so on. The NIOS continuously poll the register bank status and start the DMA transfer to the MAC accordingly.

*C. Module Management controller*

The Module Management Controller is implemented with the VT026 provided by VADATECH. The CPU is an NXP 32-bit RISC processor (LPC2138, packaged in a 10mm x 10mm, 64 pin HVQFN) with integrated flash, dual I2C, dual SPI, dual RS-232, WDT, RTC, 10-bit A/D, etc. The VT026 firmware is field upgradeable and configurable via the I2C bus. The LPC2138's can be purchased from VadaTech pre-programed. In order to enable the AMC board some minimal FRU and SDR parameters, such as the power budget or the port configuration, have to be set. This is done by using the ipmitool through an Ethernet connection with the MCH. For the power supply, a DC-DC converter is used to generate a 3.3 V voltage from the 12 V payload voltage on the backplane. The other voltages are generated, from the 3.3 Volts, using classical voltage regulators. Fig. 7 shows the board layout.

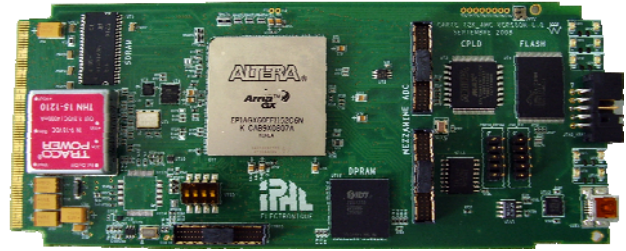

Fig. 7. ARIA-GX AMC board

## V. SYNCHRONIZATION

Obviously, a common ADC clock is mandatory on all AMC. Moreover for a soft-trigger data acquisition scheme, a distributed time is also needed to attach a timestamp to each event. IEEE1588, also called Precision Timing Protocol (PTP) is a rather recent standard [4] that defines a new method to distribute time over a Network (Ethernet or others) with an accuracy of about 1μS. Some implementations can achieve about few tens nanoseconds. Dedicated frames are exchanged between the clock master and slave nodes. The main difference with the others network timing protocols is that accurate time stamp of each frame is registered using a dedicated hardware. This means that PTP cannot be implemented using standards network material. Even if PTP defines the way to exchange accurate timestamps between master and slaves nodes, it does not define how to discipline the slave clock. A servo algorithm [6] needs to be implemented into the slave in order to regulate the slave clock drift and offset. This regulation is even more complicated when some switches are inserted in the network. PTP version 2 has been defined for this purpose and allows the measurement of the transition time of a frame into the switch. A lot of products are available in 100 Mbps, but it is not the case for Gigabit Ethernet. Nevertheless, Gigabit offers some interesting featured for PTP that simplify the servo [7]. In this paper we describe a Gigabit implementation in a FPGA, using a NIOS II μP and Gigabit MAC IP.

*A. PTP implementation with synchronous Gigabit Ethernet*

The idea is the used of the particularity of Gigabit Ethernet transceivers that continuously maintain the clock when a point to point link is established. The SLAVE node is then able to recover the clock and is directly synchronized with the MASTER. By coupling this clock scheme with a PTP implementation we can verify that the computed drift is null and that only the clock offset has to be corrected. The Block diagram is depicted on fig. 8. To emulate the PTP master clock we used a high precision 10MHz oscillator (TEMEX DOC4842) with 0.2 ppb frequency stability versus temperature. The slave and master are implemented using two ALTERA STRATIX II development boards. A PLL inside the FPGA generates the 25 MHz clock for the PHY. It replaces the 25 MHz oscillator circuit initially mounted on the board.

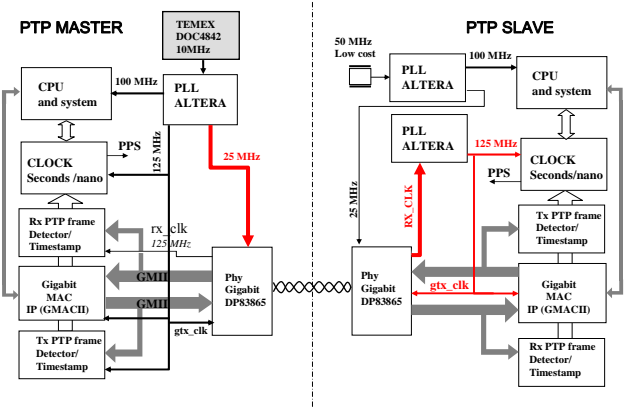

Fig. 8. PTP implementation with NIOS II and GMACII.

The PLL also generates a 125 MHz clock for the Gigabit MAC and a 100 MHz clock for the NIOS µP. The gigabit PHY is the DP83865 from National Semiconductor. The Gigabit MAC is the GMACII IP. A NIOS µP is used to control the whole system. A PTP frame detector and identifier have been implemented in the FPGA. This module polled the GMII bus in order to detect the SFD pattern (Start frame delimiter). A trigger signal is then generated in order to store a timestamp. The rest of the PTP frame is read to extract the UDP port and the frame identifier. These parameters are used to validate the PTP frame. If the frame is not a PTP, the timestamp is cancelled otherwise, the timestamp is kept and will be used by the PTP software to compute the drift and offset.

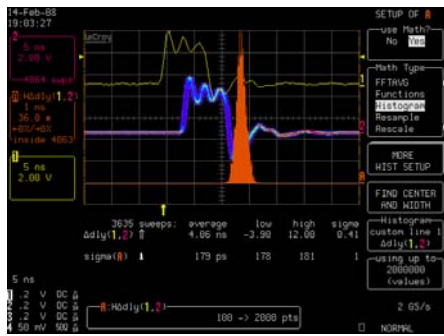

Fig. 9. Master vs. Slave PPS delay distribution.

Even if the drift can be computed using the PTP timestamps we just check that the drift is null because the clock are directly synchronized. Only the offset is corrected. Fig 9 shows the PPS delay distribution between master and slave other few hours. We observed a delay distribution of about 180 ps only due to the PLL jitter. In this implementation the clock resolution is 8 ns and the offset can not be corrected below this resolution by using the PTP itself. To further reduce the residual offset, a clock shift mechanism not yet implemented must be added at the SLAVE and controlled by the MASTER. The PTP timestamp can be used to tune the shift value in order to cancel the offset. The second main issue of this implementation is the need of specific gigabit Ethernet switch able to distribute and to propagate a synchronized clock.

### B. Synchronous Gigabit Ethernet PTP Switch in µTCA

The White Rabbit project started at CERN will fulfil these requirements [8]. They will propose a gigabit switch in the µTCA form factor based on the synchronous Ethernet that is able to propagate a clock through the gigabit link and also to compensate the offset by a clock shifting mechanism. Fig. [10]

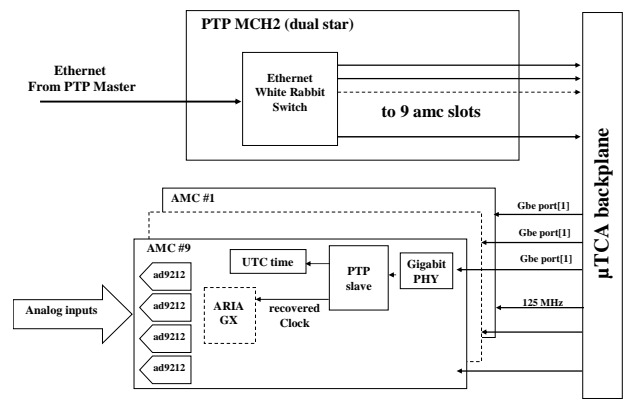

Fig. 10. PTP implementation in µTCA

The next step of our development will be to implement the WHITE RABBIT switch in our µTCA acquisition system. This will imply to implement the switch slave module inside the AMC FPGA. The fabric clock A (FCLKA) will be used to transmit the 125 MHz reference clock to each AMC. The Fabric D lanes will be used to transmit the time code. The dual start topology of the µTCA backplane will allow the separation of the timing network from the acquisition.

## VI. CONCLUSIONS

An Advanced Mezzanine Board has been realized and tested in a µTCA shelf. The data of 32 ADC channels are sent through the Gigabit Ethernet µTCA backplane. The UDP protocol is implemented in FPGA, using a Hardware acceleration architecture based on the GMACII IP and the NIOS II. The next step will consist in developing the software able to manage the full system including 12 AMC and testing in a real detector environment.

ACKNOWLEDGMENT

We thank Alain Gagnaire from ECRIN system, Saeed karamooz and Darrel Webb from VADATECH for their supports in starting up our AMC board.